\documentclass[]{spie}  

\usepackage{amsmath,amsfonts,amssymb}
\usepackage{graphicx}
\usepackage[colorlinks=true, allcolors=blue]{hyperref}

\title{Overview of the Dark Energy Spectroscopic Instrument}

\author[a,b]{Paul Martini}
\author[c]{Stephen Bailey}
\author[c,d]{Robert W. Besuner}
\author[e]{David Brooks}
\author[e]{Peter Doel}
\author[d]{Jerry Edelstein}
\author[f]{Daniel Eisenstein}
\author[g]{Brenna Flaugher}
\author[g]{Gaston Gutierrez}
\author[c]{Stewart E. Harris}
\author[b,h]{Klaus Honscheid}
\author[c,d]{Patrick Jelinsky}
\author[i]{Richard Joyce}
\author[g]{Stephen Kent}
\author[b]{Michael Levi}
\author[j]{Francisco Prada}
\author[d]{Claire Poppett}
\author[k]{David Rabinowitz}
\author[l]{Constance Rockosi}
\author[m]{Laia Cardiel Sas}
\author[b]{David J. Schlegel}
\author[n]{Michael Schubnell}
\author[o]{Ray Sharples}
\author[c]{Joseph H. Silber}
\author[i]{David Sprayberry}
\author[p,q]{Risa Wechsler}
\affil[a]{Department of Astronomy, The Ohio State University, 140 West 18th Avenue, Columbus, OH 43210, USA}
\affil[b]{Center for Cosmology and Astroparticle Physics, The Ohio State University, 140 West 18th Avenue, Columbus, OH 43210, USA}
\affil[c]{Lawrence Berkeley National Laboratory, 1 Cyclotron Road, Berkeley, CA 94720, USA}
\affil[d]{Space Sciences Laboratory, University of California, Berkeley, 7 Gauss Way, Berkeley, CA  94720, USA}
\affil[e]{Department of Physics \& Astronomy, University College London, Gower Street, London, WC1E 6BT, UK}
\affil[f]{Harvard-Smithsonian Center for Astrophysics, Harvard University, 60 Garden Street, Cambridge, MA 02138, USA}
\affil[g]{Fermi National Accelerator Laboratory, PO Box 500, Batavia, IL 60510, USA}
\affil[h]{Department of Physics, The Ohio State University, 191 West Woodruff Avenue, Columbus, OH 43210, USA}
\affil[i]{National Optical Astronomy Observatory, 950 North Cherry Street, Tucson, AZ 85719, USA}
\affil[j]{Instituto de Astrofisica de Andaluc\'{i}a, Glorieta de la Astronom\'{i}a, s/n, E-18008 Granada, Spain}
\affil[k]{Physics Department, Yale University, P.O. Box 208120, New Haven, CT 06511, USA}
\affil[l]{Department of Astronomy and Astrophysics, University of California and University of California Observatories, 1156 High Street, Santa Cruz, CA 95064, USA}
\affil[m]{Institut de Fisica d'Altes Energies (IFAE), The Barcelona Institute of Science and Technology, Campus UAB, 08193 Bellaterra Barcelona, Spain}
\affil[n]{Physics Department, University of Michigan Ann Arbor, MI 48109, USA}
\affil[o]{Institute for Computational Cosmology, Department of Physics, Durham University, South Road, Durham DH1 3LE, UK}
\affil[p]{Kavli Institute for Particle Astrophysics and Cosmology and SLAC National Accelerator Laboratory, Menlo Park, CA 94305, USA}
\affil[q]{Physics Department, Stanford University, Stanford, CA 93405, USA}

\authorinfo{Further author information: (Send correspondence to P.M.)\\P.M.: E-mail: martini.10@osu.edu, Telephone: 1 614 292 8632}

\begin{document} 
\maketitle

\begin{abstract}
The Dark Energy Spectroscopic Instrument (DESI) is under construction to measure the expansion history of the Universe using the Baryon Acoustic Oscillation technique.  The spectra of 35 million galaxies and quasars over 14000 square degrees will be measured during the life of the experiment.  A new prime focus corrector for the KPNO Mayall telescope will deliver light to 5000 fiber optic positioners.  The fibers in turn feed ten broad-band spectrographs. We present an overview of the instrumentation, the main technical requirements and challenges, and the current status of the project.
\end{abstract}

\keywords{Multi-Object Spectroscopy, Fiber Spectroscopy, Fiber Positioners, Prime Focus, Mayall Telescope, Cosmic Acceleration, Dark Energy}

\section{INTRODUCTION}
\label{sec:intro}  

The Dark Energy Survey Instrument (DESI) is an ambitous project to measure the expansion history of the universe with unprecedented precision. Over the course of a five-year survey, DESI will measure the spectra of 35 million galaxies and quasars from the present to beyond redshift three and use the Baryon Acoustic Oscillation (BAO) technique to derive cosmological parameters. The precision of the expected measurement of the dark energy equation of state is such that DESI meets the criteria for a Stage IV dark energy experiment. 

This rich dataset will also enable new insights into many other aspects of fundamental physics. Measurements of redshift space distortions will be used to search for modifications to general relativity on large scales. DESI will also measure the power spectrum of matter fluctuations with much greater precision, and over a broader range of redshifts, than previous surveys. On small scales these measurements will produce substantially improved constraints on the sum of neutrino masses. On large scales the power spectrum spectral index and its running will provide new information about inflation in the early universe. 

\begin{figure} [ht]
   \begin{center}
   \includegraphics[width=6.5in]{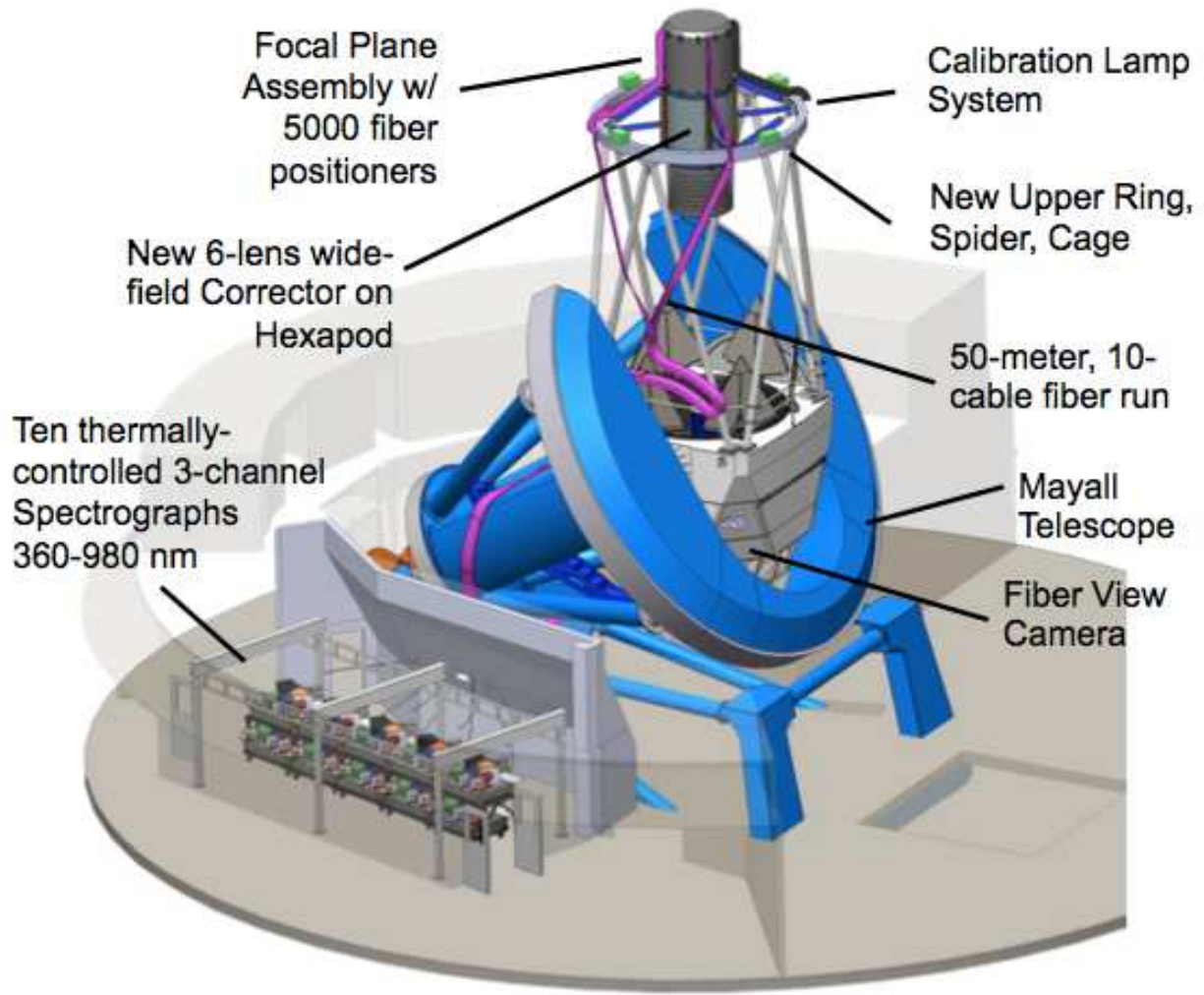}
   \end{center}
   \caption[Mayall] 
   { \label{fig:mayall} 
Model of the Mayall telescope with the addition of DESI hardware. The calibration lamp system is mounted to the new upper ring. The ring, spider, and cage supports the new corrector via a hexapod. The Focal Plane System is mounted to the end of the Corrector. Ten fiber cables, each with over 500 fibers, run from the Focal Plane System, along the telescope structure, and into a thermally-controlled room within the Mayall Large Coud\'e Room. A slithead at the end of each fiber cable directs light into a spectrograph, which disperses the light onto three detectors per spectrum. All of the instrument is controlled with the Instrument Control System. The data are rapidly transferred offsite where they are processed through a custom pipeline.}
   \end{figure}

DESI will accomplish these goals with new, highly efficient instrumentation that can measure 5000 spectra in a single observation, and rapidly reconfigure between observations. Light from the 4-m Mayall primary mirror will enter DESI through a new, prime-focus corrector with an 8 deg$^2$ field of view. There are 5000 robotic fiber positioners at the 0.8-m diameter aspheric focal surface of the corrector. Such a large number of positioners is only possible because the positioners have a minimum separation of only 10.4 mm. These positioners are part of the focal plane system, which equally divides them into 10 petals. The 500 fibers connected to each petal transmit the flux through $\sim 50$ m of cables to a slit head in one of the ten, identical spectrographs. A pair of dichroics in each spectrograph splits the light into three channels that together produce a continuous spectrum for each object from 360 -– 980 nm. Each channel has a distinct spectral resolution that ranges from 2000 -– 3000 in the shortest-wavelength channel to 4000 –- 5000 in the longest-wavelength channel. Figure~\ref{fig:mayall} shows a model of the Mayall telescope with many of the DESI hardware components. 

\section{MAYALL TELESCOPE}

The National Optical Astronomy Observatory (NOAO) has substantially improved the Mayall telescope control system (TCS) over the last few years to prepare for the DESI survey. NOAO will also replace the upper ring of the telescope and install a new set of spider vanes and a cage to support the new corrector and focal plane system. The Mayall telescope was temporarily shut down in February 2018 to begin preparations for installation of DESI. 

\subsection{Telescope System}

Most updates of the Mayall TCS are similar to the significant upgrade of the 4-m Blanco telescope for the Dark Energy Survey \cite{warner12}. The most significant physical changes are the replacement of the position servo, the addition of a velocity servo, and new hour angle and declination encoders\cite{sprayberry16}. NOAO also wrote new software for telescope control\cite{abareshi16}. The new system has been in operation for well over a year, and was used to conduct the Mayall $z$-band Legacy Survey. The pointing errors of the new system are about $3''$ RMS, which is more than a factor of six improvement relative to the previous system. 

\subsection{Cage and Ring}

The installation of DESI will include the complete replacement of the top end of the telescope, as well as routing fiber cables from the focal plane system to the spectrographs\cite{sprayberry14}. The top end replacement is because the Mayall telescope was built to support both an $f/3$ prime focus and an $f/8$ Cassegrain focus. The top end was consequently designed as a pair of concentric rings with a motor that could flip between these two foci. The new top end for DESI will include a substantially lighter single ring, as well as new vanes and a cage that connects to the hexapod. The hexapod supports the corrector barrel, and the focal plane system is bolted to the corrector barrel. The much lower weight of the new top ring effectively compensates for the greater weight of the new corrector and focal plane system such that the net change to the telescope mass and moment is fairly modest. The latest status of work at the Mayall is presented elsewhere in these proceedings\cite{allen18}.  

\section{CORRECTOR}

   \begin{figure} [ht]
   \begin{center}
   \includegraphics[width=6.5in]{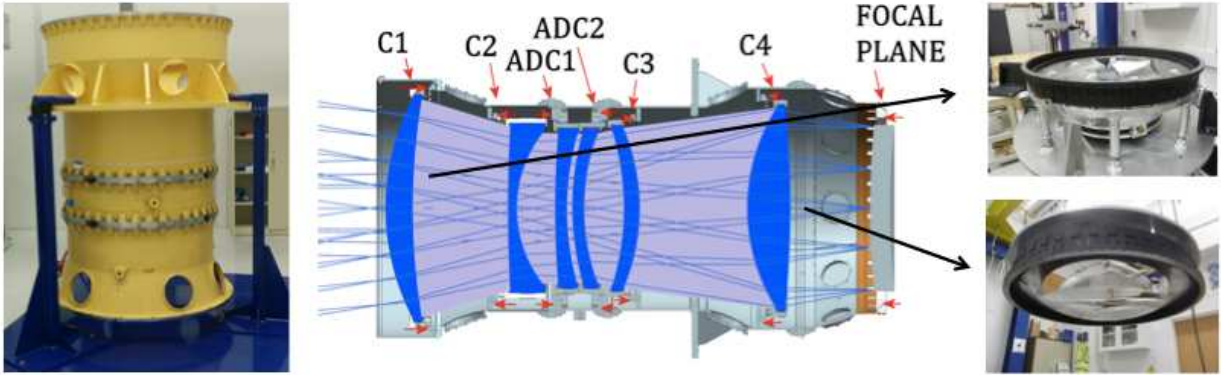}
   \end{center}
   \caption[Corrector] 
   { \label{fig:corrector} 
Elements of the new corrector. The leftmost photo shows the completed corrector barrel at Fermilab before it was painted and shipped to University College London. The middle panel shows a model of the optical design and mechanical support structure. The two photos on the right show lens C1 (top) and C4 (bottom) mounted in their cells. The complete corrector is scheduled to ship to the Mayall in Summer 2018.}
   \end{figure} 
   
The DESI prime focus corrector has a $3.2^\circ$ diameter field of view with six 0.8 to 1.14-m diameter lenses (see Figure~\ref{fig:corrector}). These lenses include a two-element Atmospheric Dispersion Compensator (ADC). The average focal ratio is $f/3.86$ and the average plate scale on the aspheric focal surface is $254.8$ mm/deg. The six lenses are mounted in cells that are held in position within a steel barrel assembly. A six degree-of-freedom hexapod maintains the alignment of the corrector barrel plus focal plane system to the telescope primary mirror. 

\subsection{Optical design}

The constraints on the optical design include operation over the wavelength band 360 -- 980 nm, good performance at zenith angles up to $60^\circ$, modest focal plate curvature, and minimal chief ray deviation\cite{doel14}. The corrector meets the requirements with all spherical surfaces, with the exception of one aspheric surface each on elements C2 and C3. The lens blanks for the all-spherical lenses C1 and C4 were produced by Corning, Inc. and polished by L3 Brashear. The lens blanks for C2 and C3 were produced by Ohara Corporation and polished by Arizona Optical Systems. One each of the ADC lens blanks were produced by Ohara Corporation and Schott North America, Inc., and Rayleigh Optical Corporation polished both lenses\cite{miller18}. Viavi Solutions, Inc. coated all six corrector elements. The performance of the coatings met the requirement of $98.5$\% average transmission per surface. The coating of the lenses is described elsewhere in these proceedings\cite{kennemore18}. All six lenses have been polished and coated, and final alignment and integration is currently in progress at University College London\cite{doel18}. The corrector is scheduled to ship to the Mayall in Summer 2018.

\subsection{Mechanical structure}

The mechanical design and construction of the corrector was led by the Fermi National Accelerator Laboratory\cite{gutierrez18}, and is based on their similar design for the DECam instrument\cite{depoy08}. The carbon steel corrector barrel has three main sections that support the optics. The Front section supports the C1 and C2 lenses, and the Aft section supports the C3 and C4 lenses. The Middle section of the barrel supports the ADC. The ADC mechanism is based on a custom bearing driven by harmonic drive motors. The hexapod attaches to the Aft section of the barrel and the cage. Its purpose is to adjust the location of the corrector and focal plane system as needed to maintain optical alignment with the telescope primary mirror. The barrel also contains a shroud or light baffle that faces the primary, and an adapter that supports the focal plane assembly.

The lenses are mounted in cells that accommodate the differences in the coefficient of thermal expansion between the lens materials and the steel barrel, as well as maintain alignment over the full range of zenith angles. The cell material is a nickel-iron alloy, and a lens is attached to each cell with axial and radial RTV pads. The thicknesses of the radial pads produce an athermal design, while the axial pads both support the lenses over the full range of gravity vectors (0-60$^\circ$ Zenith angle) and account for surface irregularities that could produce uneven loading of the lenses. The lateral displacement tolerances for the cells are $\pm$20$\mu$m and the tip/tilt tolerances are about 20$\mu$m at the edge of the lenses. The steel barrel is presently at University College London for the integration of the optical elements. 

\section{FOCAL PLANE SYSTEM}

\begin{figure} [ht]
   \begin{center}
   \includegraphics[width=6.5in]{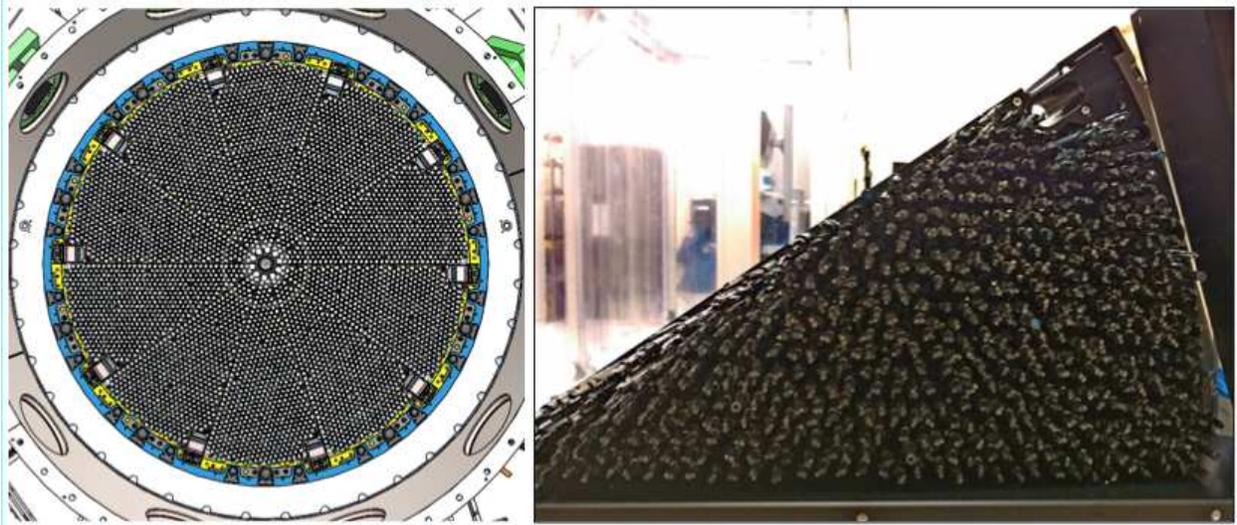}
   \end{center}
   \caption[FPS] 
   { \label{fig:fps} 
The focal plane system includes 5000 robotic fiber positioners, 6 guide cameras, 4 wavefront cameras, and 120 illuminated fiducials. These are mounted on ten, identical petals, and the fibers from each petal connect to one spectrograph. The image on the left shows a model of the focal plane system as viewed from the corrector, while the photo on the right shows the first, fully populated petal at LBNL.}
   \end{figure}

The focal plane system includes 5000 robotic fiber positioners, 6 guide cameras, 4 wavefront cameras, and 120 illuminated fiducials\cite{silber18} (see Figure~\ref{fig:fps}). These elements are in a thermal enclosure that minimizes the addition of waste heat into the dome environment. There is also a fiber view camera mounted at the primary mirror that provides feedback on fiber positions. The focal plane system is divided into ten identical “petals” that are $36^\circ$ wedges that fill the circular focal plane. Each petal has 500 fibers that feed a single spectrograph, two additional fibers that connect to a sky monitor camera, 12 illuminated fiducials, and a Guide/Focus/Alignment camera (GFA). The design and integration of the focal plane system is led by Lawrence Berkeley National Lab (LBNL), with contributions from many partner institutions. The scheduled delivery date to the Mayall is March 2019. 

\subsection{Fiber Positioners}

The fiber positioner robots have two rotational axes that can place the fiber within a 12-mm diameter patrol region. The fiber positioners have a 10.4-mm pitch between neighboring units. The close proximity of the positioners is a critical feature that enables the extraordinary multiplex of DESI. The moderate potential overlap between neighboring fiber positioners requires consideration of collisions as part of both target selection and the the orchestration of positioner moves to their target positions. Fiber assignment software identifies feasible targets for each observation and anticollision software plans a safe sequence of acquisition moves for that set of targets. 

The fiber positioners are required to place the fiber tips within $\leq 5\mu$m RMS of their targets. As the fiber positioners are less precise for large moves, we achieve this requirement with an iterative process. After each move  of the positioners, the fiber tips and fiducials are illuminated and imaged with the fiber view camera. Software measures the location of each fiber relative to the target location,  determines any correction moves that may be necessary, and then sends these correction moves to the fiber positioners. Lab tests indicate we will achieve 1 -- $2\mu$m RMS position accuracy with one or two correction moves\cite{schubnell18}. The University of Michigan is responsible for production of the fiber positioners\cite{aguilar18}. At present more than 90\% are complete and the remainder are expected in Summer 2018. 

\subsection{Guiding and wavefront sensing}

The GFA cameras are located at the periphery of the focal plane. Six of the cameras will be used for guiding. The remaining four will be used for wavefront sensing with intra- and extra-focal images. All of the GFAs will have an $r$-band filter that does not pass the bluer light used by the fiber illumination system. The guide sensors have a flat, 5-mm thick filter. Each wavefront sensor will have half of the active area covered by a 1.625-mm thick filter, and the other half covered by an 8.375-mm thick filter. These filters will place intra- and extra-focal images on each wavefront sensor. The GFAs are under construction by the Institut de Fisica d'Altes Energies (IFAE) and the Institut de Ciencies de l'Espai (ICE CSIC, IEEC). 

The Active Optics System (AOS) uses the wavefront sensor images to determine if there are wavefront errors that represent a misalignment of the optical system, and provides correction moves to the hexapod. The AOS is based on the similar system developed for the DECam instrument\cite{roodman14}. In addition to the different optical designs of DECam and DESI, the DESI wavefront sensors are not behind the instrument shutter and therefore can compute corrections before the start of an exposure. 

\subsection{Fiber View Camera}

The purpose of the fiber view camera system is to rapidly measure the locations of the fiber positioners. The sensor is a Finger Lakes Instrumentation Microline camera body with Kodak KAF-50100 CCD mounted to a 600 mm f/4 Canon lens. The CCD has 6132x8176 pixels, a pixel pitch of $6\mu$m, and a three second readout time.  The camera observes the focal surface from a distance of 12.25 m, about 1 m behind the primary mirror.   
The software takes images of the illuminated fiducials and fiber tips and returns positions with $\leq3\mu$m RMS error within three seconds. Yale University developed both the fiber view camera and the illuminated fiducials that provide a static reference for the fiber positioners. This system was tested at the Mayall in 2016 as part of the ProtoDESI experiment\cite{fagrelius18}. 

\section{FIBER SYSTEM}

The fiber system extends from the fiber positioners to the slit heads in the spectrographs. The three main components of the fiber system are the Positioner Fiber Assembly (PFA), the fiber cable, and the slit head. The PFAs are mounted into the fiber positioners. These are fusion spliced to the 47.5-m fiber cables, and finally terminate in slit blocks that make up the slit head in each spectrograph. Durham University is responsible for the production of the fiber cables, and LBNL for integration of the fiber cables with the focal plane system. The fiber system is more fully described elsewhere in these proceedings\cite{poppett18}.  

\subsection{Positioner Fiber Assemblies}

Light enters the fiber system through the positioner fiber assemblies, which are bonded into the fiber positioners. Each of these assemblies is a precision-cleaved, $\sim$3.1-m length of fiber with an anti-reflection coating that is bonded to a fused silica ferrule and then bonded into a positioner. The other end terminates in a fiber spool box that provides strain relief. Each positioner fiber assembly is fusion-spliced to the fiber cable at LBNL, which produces a negligible impact on throughput. 

\subsection{Fiber Cable and Slitheads}

There are ten fiber cables, each with over 500 fibers. Each cable includes 6\% more spare fiber to allow for damage. The center of the cable has a tensile element that limits stresses due to axial loads or temperature changes. The fibers are spiral-wound to avoid cumulative tension. The cables run across the support vanes for the prime focus corrector, down the telescope truss, and into the Large Coud\'e Room. The cable guides along the support structure limit cable twists, and allow a bend radius no less than 200 mm. 

There is one slithead per spectrograph. Each slithead has 500 fibers arranged into 20 blocks of 25 fibers each. Each block has 25 parallel V-grooves, and each block is tilted so that the slit matches the convex focal surface of the spectrograph. 

\section{Spectrographs}

\begin{figure} [ht]
   \begin{center}
   \includegraphics[width=6.5in]{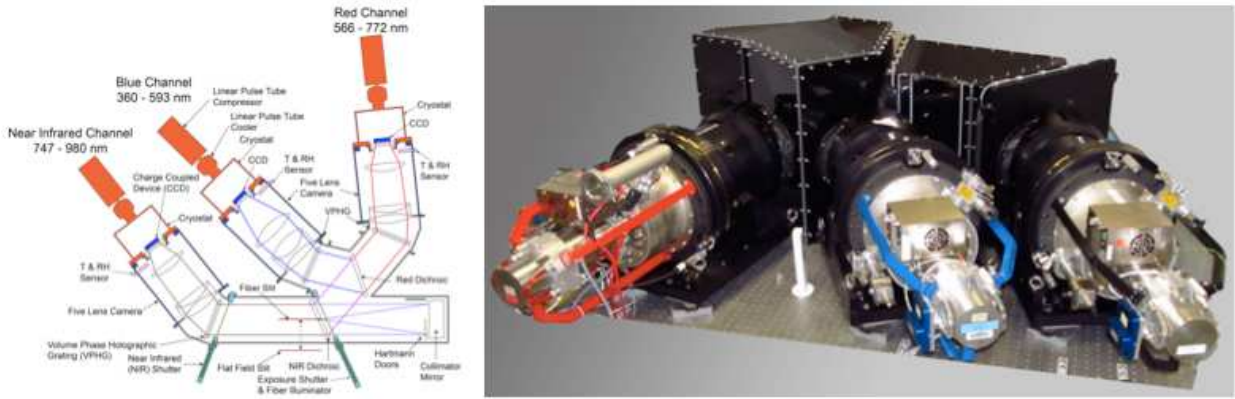}
   \end{center}
   \caption[Specs] 
   { \label{fig:spec} 
Optical design of the DESI spectrographs (left) and a photo of the first spectrograph (right). Light enters the spectrographs through the slithead, which is behind the exposure shutter. It reflects off the collimator mirror, and then is split by two dichroics into three wavelength channels.}
   \end{figure}

DESI has ten identical spectrographs. Each spectrograph has three wavelength channels (blue, red, near-infrared) in order to optimize throughput and spectral coverage, and each channel has a distinct resolution range. Figure~\ref{fig:spec} shows an overview of the spectrographs. The spectrographs are under construction at Winlight Optical Systems in France with contributions from many partners\cite{edelstein18}. The first spectrograph has shipped to the Mayall telescope, and the remaining nine are scheduled to arrive through Spring 2019.  

\subsection{Optical design}

The light from the slithead passes through a slot in the near-infrared (NIR) dichroic and reflects off a $f/3.57$ spherical collimator mirror with an enhanced silver coating. Light for the NIR channel passes through the NIR dichroic, is dispersed by the NIR volume phase holographic (VPH) grating, and enters the NIR camera. The light for the blue and red channels reflects off the NIR dichroic and the blue light reflects a second time off the red dichroic, is dispersed by the Blue VPH grating, and enters the blue camera. The red light passes through the red dichroic, through the red VPH grating, and into the red camera. The resolutions of the blue, red, and near-infrared spectra are approximately 2000 -- 3000, 3500 -- 4500, and 4000 -- 5500, respectively. 

Materion produced the dichroics and Kaiser Optical Systems, Inc. produced the VPH gratings. Winlight Optical Systems produced the lenses and collimator mirrors. Winlight is also responsible for building the cameras and the alignment of the spectrographs. The blue cameras use a fluid coupled design that Winlight adapted from the K/COSMOS instruments\cite{obrien14}. The other two cameras have bonded optical elements. Personnel from Laboratoire de Physique Nucl\'eaire et des Hautes \'Energies test each spectrograph at Winlight Optical Systems once alignment is complete\cite{perruchot18}.  

\subsection{Detectors and electronics}

The three CCDs in each spectrograph have 4096x4096 pixels with a $15\mu$m pitch and a requirement of $<3$e$^-$ readout noise. They are read out through four amplifiers at a pixel rate of 100kHz. The blue channel detector is an STA device packaged by the University of Arizona Imaging Technology Laboratory. The requirements include quantum efficiency $>75$\% from 360 -- 400 nm and $>85$\% from 400 -- 600 nm. LBNL produced the red and NIR devices. Their requirements include quantum efficiency $> 85$\% from 600 -- 900 nm and $>60$\% from 900 -- 980 nm. LBNL designed and built the detector front end electronics. 

\subsection{Cryostats}

The Commissariat \`a l'\'energie atomique Saclay is responsible for the design and fabrication of the detector cryostats. The requirements on the cryostats include precise $\pm1$K and stable $\pm0.1$K temperature control, and they must accommodate the $\sim$170K and $\sim$140K operational temperatures of the STA and LBL devices, respectively. The cryostats maintain temperature control with linear pulse tube cryocoolers. The cryostats are delivered to Winlight as part of the integration of the spectrographs, and are then shipped to the Mayall with the spectrographs. 

\subsection{Mechanisms}

Each spectrograph has two shutters and a pair of Hartmann doors. The exposure shutter is immediately adjacent to the NIR dichroic and slithead and its operation sets the length of individual exposures. The exposure shutter has the fiber illumination system integrated into the shutter blade\cite{derwent16}. This system includes a strip of LEDs that is mounted behind a diffuser, a blue filter, and a slit. Because the fiber illumination system is very bright, and will normally be used during readout of the previous exposure, the shutter has an inflatable seal that prevents this light from reaching the blue and red channels. There is a second, NIR shutter with an inflatable seal that is mounted in front of the NIR grating that is closed before operation of the fiber illumination system. Together, these two shutters and baffles form a light-tight enclosure during operation of the fiber illumination system. The two Hartmann doors are attached to the collimator mount, and are used for some optical alignment and focus tasks. The shutters, fiber illumination system, collimator mounts with Hartmann doors, mechanism electronics, and dichroic mounts were designed and build by The Ohio State University. 

\subsection{Thermal enclosure and support}

The spectrographs are located in the Large Coud\'e Room of the Mayall telescope. The spectrographs were designed for optimal performance at a temperature of $20 \pm 2^\circ$ C. They also have requirements on maximum humidity, environmental cleanliness when the optics are exposed, and various service and access requirements. The enclosure will be a 15.5$\times$35 foot insulated room with a Class 10,000 rating. Western Environmental Corporation has provided the parts for the enclosure, and NOAO personnel will assemble the enclosure in Summer 2018. The enclosure will include five upper and five lower racks built by Ohio State that house the ten spectrographs, as well as an overhead gantry crane for service functions. There is also a separate ``Annex'' space that will be used for spectrograph re-assembly and any extensive repair work that may be necessary during operations. 

\section{SOFTWARE}

The DESI software effort involves substantial development in multiple areas. The Instrument Control System operates the hardware, provides the observer interface, and controls the communication between various subsystems. The Data Systems software includes the target selection pipeline, survey strategy, spectroscopic pipeline, data transfer, distribution, and archiving. 

\subsection{Instrument Control System}

The Instrument Control System\ (ICS) is essentially the nervous system for the entire project, and the development of this system was led by The Ohio State University. The ICS connects to every networked device and software system and controls the operation of every aspect of the instrument. The observation console provides the user interface to the instrument, and operates the TCS, spectrographs, cryostats, the fiber view camera, fiber positioners, and the GFAs. The ICS also reads data from the spectrographs and images from the GFAs and FVC. The spectroscopic data are then analyzed by data systems, the GFA data are used to compute tracking corrections for the TCS or wavefront corrections for the hexapod, and output fiber positions from the FVC are used to determine the correction moves for target acquisition. The computer hardware for the ICS has been installed at the Mayall and the system is ready to support the acceptance of the other hardware components as they arrive\cite{honscheid18}.

\subsection{Data Systems}

DESI target selection uses deep $g-$, $r-$, and $z-$band observations of the 14,000 deg$^2$ DESI footprint by the Legacy Surveys\cite{dey18} combined with all-sky observations at $3.4$ and $4.6\mu$m from the Wide-field Infrared Survey Explorer (WISE).  The Legacy Surveys are public surveys based on three imaging projects at different telescopes: The Beijing-Arizona Sky Survey (BASS), The DECam Legacy Survey (DECaLS), and The Mayall z-band Legacy Survey (MzLS). 

Survey strategy software optimizes field selection and fiber assignment within fields. During operations, a sky camera will use multiple measurements of the sky continuum brightness to dynamically adjust the exposure times and produce consistent signal-to-noise ratio data. The sky camera uses dedicated fibers from the focal plane system that do not connect to the spectrographs and is under construction by the University of California at Irvine. All data are transferred to the National Energy Resarch Scientific Computing Center (NERSC). Tests of the data pipeline have shown that data can be transferred to NERSC within minutes and a complete night of data can be processed in about six hours. The data products will be immediately available to the collaboration. The development of these software components is led by LBNL.  

\section{Commissioning}

Commissioning includes all on-sky tests with the instrument before the start of the survey. The main categories of commissioning tasks are the TCS, guiding, the Active Optics System, fiber positioning, fiber spectroscopy, and software. The project has allocated six months to complete these tasks, which start after the installation of the focal plane system. The project schedule includes several months between the installation of the corrector and the arrival of the focal plane system. The project is consequently building a separate Commissioning Instrument based on five commercial CCD cameras that can complete a substantial number of the first three categories of commissioning tasks. The addition of the Commissioning Instrument does not impact the project schedule, will decrease the time required for commissioning of the focal plane system, and will enable the early retirement of a number of risks related to commissioning activities. 

\subsection{Commissioning instrument}

The Commissioning Instrument has five commercial SBIG STXL-6303 cameras and 22 illuminated fiducials that will be precisely located on the aspheric focal surface of the corrector. These cameras and fiducials are supported by a large steel structure that matches the expected mass and moment of the focal plane system. Four of the five commercial cameras are located at the same off-axis distance as the GFAs on the focal plane system, while the fifth will provide unique, on-axis imaging through the corrector. Direct imaging with the five cameras will be used to commission the TCS and guider algorithm, and the telescope will be defocused to produce intra- or extra-focal images to start to commission the Active Optics System. An overview of the Commissioning Instrument is presented elsewhere in these proceedings \cite{ross18}. 

Astrometric solutions for the five cameras, combined with images of the 22 illuminated fiducials with the fiber view camera, will be used to measure the optical distortions of the corrector. This measurement is critical for the accurate alignment of fiber positioners with the focal plane system. This solution requires $5\mu$m and $10\mu$m RMS measurement uncertainties between the CCD pixels and the nearest fiducial, and between fiducials, respectively. This metrology program is described elsewhere in these proceedings\cite{coles18}. 

\subsection{Focal Plane System}

Commissioning of the focal plane system will begin several months after the end of observations with the Commissioning Instrument. This time will be used to install the focal plane system, run the fiber cables through the telescope structure, and position the slitheads in the spectrographs. We expect to begin commissioning with a subset of the spectrographs, and then integrate the remainder as they arrive at the Mayall during the beginning of the commissioning period. Commissioning will start with a repeat of the TCS, guiding, and AOS tasks, as we expect some changes to various parameters. Several months of commissioning time is then scheduled to optimize fiber positioning and spectroscopy. Some expected challenges are thermal expansion and contraction of the focal plane system due to changes in the ambient temperature, and the transformations between target celestial positions, fiber positioner coordinates, and the fiber view camera. A substantial amount of time is also planned to minimize the inter-exposure time, and to obtain sufficiently deep exposures of the target classes to fully exercise the spectroscopic pipeline and provide feedback on target selection. 

\section{SUMMARY}

The Dark Energy Spectroscopic Instrument is more than halfway through the construction phase. A number of components have already arrived at the Mayall, and a large number will be delivered before the end of 2018. The Mayall telescope is presently in the midst of major changes to the top end. The corrector barrel and optics are complete, and will be integrated and then shipped to the Mayall in Summer 2018. On-sky observations with the Commissioning Instrument will begin in Autumn 2018. The first spectrograph has arrived, and the remaining nine will arrive through Summer 2019. The focal plane system is scheduled to arrive in March 2019. Commissioning is scheduled to begin by Summer 2019, followed by the start of the survey around the end of that year. 

\acknowledgements

This research is supported by the Director, Office of Science, Office of High Energy Physics of the U.S. Department of Energy under Contract No. DE–AC02–05CH1123, and by the National Energy Research Scientific Computing Center, a DOE Office of Science User Facility under the same contract; additional support for DESI is provided by the U.S. National Science Foundation, Division of Astronomical Sciences under Contract No. AST-0950945 to the National Optical Astronomy Observatory; the Science and Technologies Facilities Council of the United Kingdom; the Gordon and Betty Moore Foundation; the Heising-Simons Foundation; the National Council of Science and Technology of Mexico, and by the DESI Member Institutions. The authors are honored to be permitted to conduct astronomical research on Iolkam Du'ag (Kitt Peak), a mountain with particular significance to the Tohono O'odham Nation.


\end{document}